\documentclass[aps,prl,floatfix,twocolumn,tightenlines,amsmath,amssymb,footinbib]{revtex4-1}

\usepackage[T1]{fontenc}
\usepackage{graphicx}
\usepackage{amssymb}
\usepackage{hyperref}
\hypersetup{
colorlinks=true,
linkcolor=red,
citecolor=red,}
\usepackage{color}
\usepackage{amsthm}
\usepackage{psfrag}
\usepackage{ifsym}
\usepackage{dsfont}
\usepackage{enumerate}
\usepackage{amsfonts}
\usepackage{multirow} 
\usepackage{amsmath}
\usepackage{soul}

\newcommand{\bra}[1] {\langle #1 |}
\newcommand{\ket}[1] {| #1 \rangle}

\newcommand{\ketbra}[1]{ | #1 \rangle\!\langle #1 |}

\newcommand{\expec}[1]{\left\langle #1 \right\rangle}
\newcommand{\one}{\leavevmode\hbox{\small1\normalsize\kern-.33em1}}

\newcommand{\be}{\begin{equation}}
\newcommand{\ee}{\end{equation}}

\newcommand{\id}{\mathds{1}}

\DeclareMathOperator{\tr}{tr}

\newcommand{\mm}{{\cal M}}

\newcommand{\Proj}[1]{[[#1]]}

\begin{document}
\title{A Unifying Framework for Spatial and Temporal Quantum Correlations}

\author{Fabio Costa$^{1\ast}$, Martin Ringbauer$^{1,2,3\ast}$, Michael E. Goggin$^{1,4}$, Andrew G. White$^{1,2}$, and Alessandro Fedrizzi$^{3}$}

\affiliation{$^{1}$Centre for Engineered Quantum Systems, $^{2}$Centre for Quantum Computer and Communication Technology, School of Mathematics and Physics, University of Queensland, Brisbane,   QLD 4072, Australia\\
$^3$Institute of Photonics and Quantum Sciences, School of Engineering and Physical Sciences, Heriot-Watt University, Edinburgh EH14 4AS, UK\\
$^4$Department of Physics, Truman State University, Kirksville, MO 63501, USA\\
$^\ast$These authors contributed equally to the manuscript}

\begin{abstract}
Measurements on a single quantum system at different times reveal rich non-classical correlations similar to those observed in spatially separated multi-partite systems. Here we introduce a theory framework that unifies the description of temporal, spatial, and spatio-temporal resources for quantum correlations. We identify, and experimentally demonstrate simple cases where an exact mapping between the domains is possible. We then identify correlation resources in arbitrary situations, where not all spatial quantum states correspond to a process and not all temporal measurements have a spatial analogue. These results provide a starting point for the systematic exploration of multi-point temporal correlations as a powerful resource for quantum information processing.
\end{abstract}

\maketitle

Quantum correlations are typically revealed when two (or more) parties perform local measurements on \emph{spatially} separated quantum systems. However, a similarly rich structure of correlations appears for \emph{temporally} separated measurements on a single quantum system. Temporal quantum correlations were first discussed by Leggett and Garg, who showed that two sequential quantum measurements on a macroscopic system can reveal correlations stronger than those allowed under the classical assumptions of macroscopic realism and measurement non-invasiveness~\cite{leggett1985qmm}. Leggett-Garg inequalities have been tested for microscopic systems \cite{palacios2010evb,goggin2011vlg,waldherr2011vtb,knee2012vlg} and more recently for a superconducting flux quantum bit approaching the required complexity for macroscopic realism~\cite{knee2016set}. 

Temporal correlations have since been identified as a resource for quantum information tasks~\cite{brukner2004qet,markiewicz2014genuinely} and continue to be explored for foundational questions~\cite{Markiewicz2014, Brierley2015} including in hybrid spatio-temporal inequalities~\cite{dressel2014alh,white2016ped}. Although they share many features with spatial correlations, there are significant differences, both qualitative, e.g. the violation of monogamy of entanglement~\cite{brukner2004qet}, and quantitative, e.g. Hardy's paradox is stronger in time~\cite{fedrizzi201hpv}. This raises the question about the precise relationship of these two scenarios.

\begin{figure}[h!]
  \begin{center}
  \includegraphics[width=0.70\columnwidth]{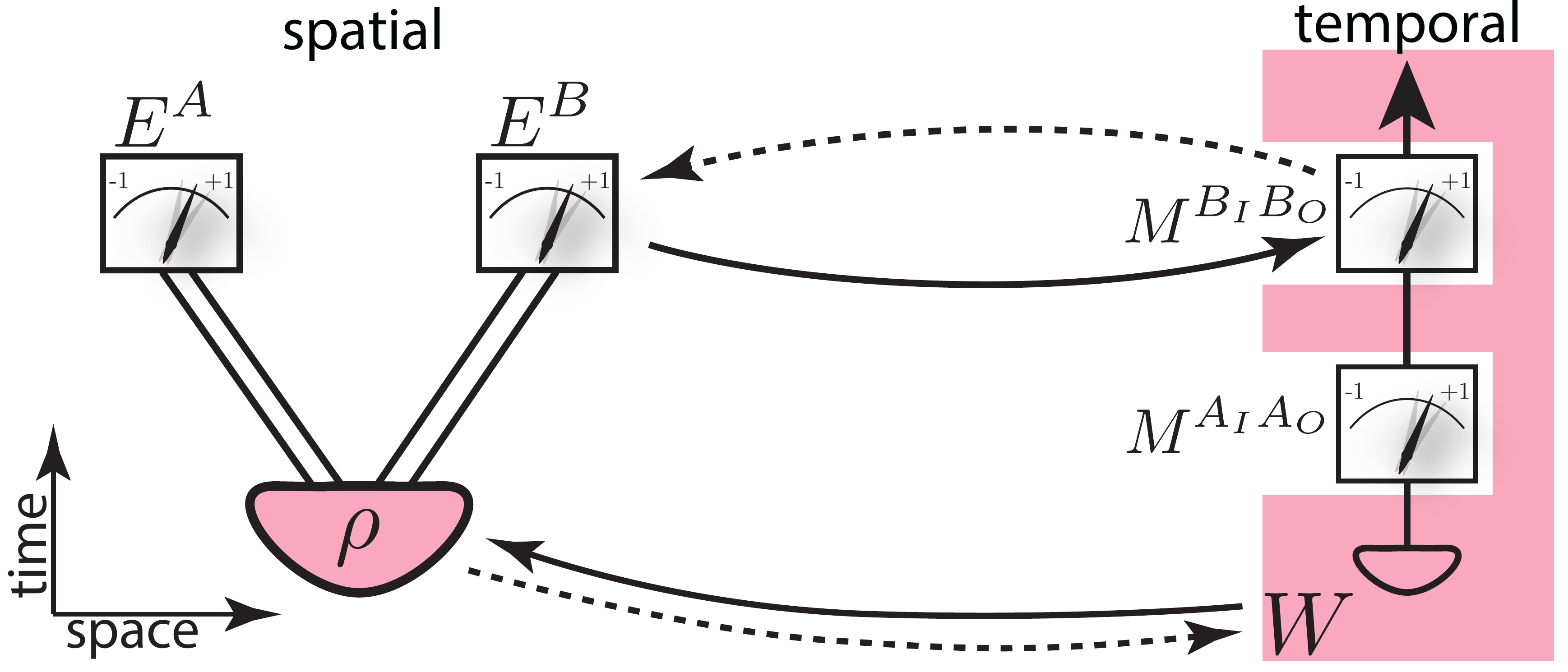}
  \end{center}
\vspace{-1.5em}
\caption{\textbf{Spatial vs.~temporal quantum correlations.} Spatial quantum correlations are revealed by local measurements $E^A$ and $E^B$ on spatially separated quantum systems, whereas temporal correlations are observed between subsequent measurements $M^{A_IA_O}$ and $M^{B_IB_O}$ on the same system. The solid arrows indicate that every spatial measurement has a temporal analogue, while the converse is not true in general (dashed arrow). Since temporal measurements must specify the post-measurement state, we compare them to spatial measurements of systems of double the dimension of the temporal counterpart (cf. the double solid lines). 
The resource (red shading) generating the observed spatial correlations is the initial quantum state $\rho$, whereas in the temporal case it is the quantum process $W$. Solid (dashed) arrows indicate that every $W$ can be mapped to a $\rho$ but not vice versa.}
  \label{fig:Motivation}
\end{figure}

Here we introduce a unified treatment of spatial and temporal resources for quantum correlations, clarifying the relationship between spatially separated quantum states, temporal quantum processes, and measurements in the two domains. We show that there are spatial quantum states that cannot be identified with temporal processes, and temporal measurements that do not have a spatial analogue, see Fig.~\ref{fig:Motivation}. We then describe the resources for quantum correlations in the two domains and discuss when a one-to-one mapping exists. We explore this experimentally in a tripartite spatio-temporal scenario that mirrors spatial Greenberger-Horne-Zeilinger-(GHZ) entanglement. Finally, we propose a general no fine-tuning criterion to understand how spatial or temporal quantum resources outperform classical resources with the same no-signalling relations. Using this criterion we rule out non-fine-tuned hidden-variable models for our experimental spatio-temporal correlations by violating a Svetlichny inequality.

\paragraph{Background---} We start with Alice and Bob, who each perform one of two measurements $X$ and $Y$ on individual quantum systems, respectively, and obtain $\pm1$-valued outcomes $A$ and $B$. For space-like separated parties, all correlations between $A$ and $B$ that can be explained by
classical cause-effect relations must satisfy the Clauser-Horne-Shimony-Holt (CHSH)~\cite{Clauser1969} inequality, 
\begin{equation}
S_{\textsc{chsh}}^{\textsc{ab}} = \expec{A_0B_0} - \expec{A_0B_1} + \expec{A_1B_0} + \expec{A_1B_1} \leq 2,
\label{eq:CHSH}
\end{equation}
where $\expec{A_xB_y}=\sum_{a,b} ab P(a,b\vert x,y)$ denotes the joint expectation value for Alice's and Bob's measurements choice $x$ and $y$, respectively. Correlations obtained from measurements on entangled quantum states can violate inequality~\eqref{eq:CHSH} up to $S_{\textsc{chsh}} = 2\sqrt{2}$~\cite{Tsirelson1980}. By imposing realism and no-signaling in time instead of realism and local causality~\cite{Fritz2010,brukner2004qet}, one can derive a CHSH inequality for temporally separated measurements on the same quantum system. A violation of this inequality, demonstrated experimentally in~\cite{fedrizzi201hpv}, indicates the presence of \emph{entanglement in time}. Its precise relationship to the usual notion of entanglement, however, remains a matter of debate.

\begin{figure}[h!]
  \begin{center}
  \includegraphics[width=0.9\columnwidth]{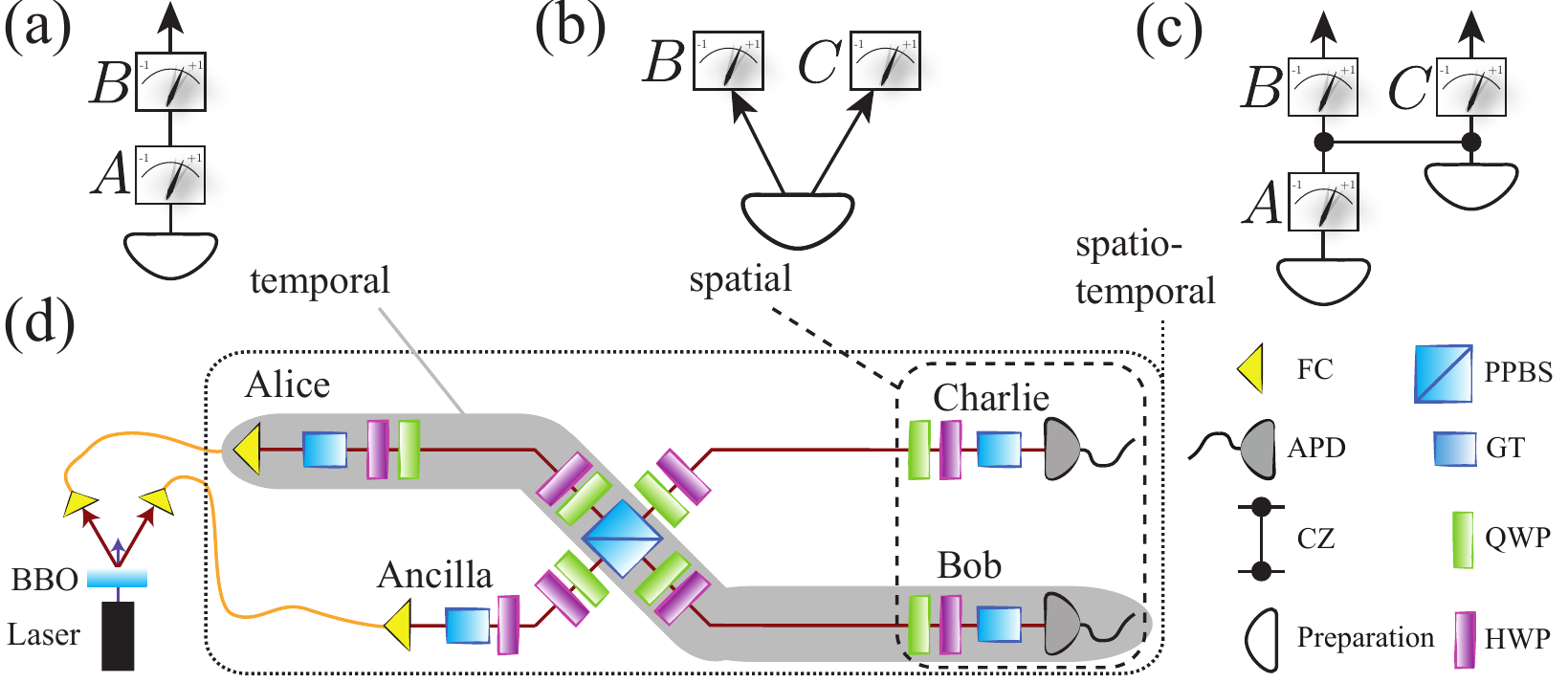}
  \end{center}
\vspace{-1.5em}
\caption{\textbf{Spatial, temporal and spatio-temporal correlations.} \textbf{(a)} In the temporal scenario, Alice and Bob measure the same quantum system at different times. \textbf{(b)} In the traditional spatial scenario, Alice and Bob perform local measurements on parts of a shared quantum system. \textbf{(c)} In a general scenario, there may be multiple parties, Alice, Bob, Charlie, etc.\ who perform temporally and spatially separate measurements in a quantum network. 
\textbf{(d)} A simple experiment with two interacting quantum systems already exhibits all three forms of correlations. Pairs of single photons are generated via spontaneous parametric downconversion in a fs-pumped $\beta$-Barium-borate (BBO) crystal. The polarization state of the system photon is prepared by Alice using a half-waveplate (HWP), quarter-waveplate (QWP) and Glan-Taylor polarizer (GT). The ancilla photon is entangled non-deterministically to the system via two-photon interference in a partially-polarizing beam splitter (PPBS)~\cite{Langford2005}. Finally Bob and Charlie measure the states of system and ancilla, using a HWP, QWP and GT, before detecting their photons with avalanche photodiodes (APD). In this experiment, Alice and Bob observe temporal correlations on the system qubit; Bob and Charlie observe spatial correlations between the final measurements; and taking into account all three parties, we observe spatio-temporal correlations (dotted line).}
  \label{fig:Setup}
\end{figure}

\paragraph{Theoretical framework---} To address this question, we now introduce a unifying framework for temporal and spatial resources for quantum correlations, where a resource is understood as a physical system or device that can be accessed to extract correlations, Fig.~\ref{fig:Setup}. There are several formally equivalent frameworks for studying temporal processes and multipartite states on an equal footing, such as quantum channels with memory~\cite{Kretschmann2005}, quantum strategies~\cite{gutoski06}, quantum combs~\cite{gutoski06, chiribella08, Chiribella2009, Bisio2011}, the two-state vector formalism~\cite{Aharonov1964, Aharonov2009, Silva2014, Silva2017}, the ``general boundary'' formalism~\cite{Oeckl2003318}, operational open dynamics~\cite{modioperational2012, pollockcomplete2015}, and quantum causal models~\cite{costa2016, Allen2016}. We will use the semantics and conventions of the ``process matrix'' formalism~\cite{oreshkov12, araujo15, oreshkov15}, since it allows a clean distinction between resources and operations.

The starting point for our framework is that a local operation in a temporal scenario is described using \emph{two} Hilbert spaces: the input space $A_I$, representing the system before the measurement, and its output space $A_O$ thereafter~\footnote{Other, related approaches attempt to describe temporal correlations using a single Hilbert per operation~\cite{Leifer2013, Fitzsimons2015}. However, it is problematic to extend these approaches to a consistent general framework~\cite{Horsman20170395}.}. A general measurement is described as an \emph{instrument}~\cite{davies70}: a collection of completely-positive (CP) trace non-increasing maps, $\{{\cal M}_{a|x}\}_a$, with ${\cal M}_{a|x}:A_I \mapsto A_O$, where $x$ labels the measurement setting and $a$ the outcome, such that $\sum_a {\cal M}_{a|x}$ is completely positive and trace preserving (CPTP). State preparations and final measurements (after which the system is discarded) are recovered as special cases with trivial (i.e., one-dimensional) input and output spaces, for which CP maps reduce to density matrices and Positive Operator Valued Measures (POVMs), respectively.

Using the Choi-Jamio{\l}kowski isomorphism~\cite{jamio72, choi75}, we describe local measurements for a party $A$ as matrices $M_{a|x}^{A_IA_O}$ on the space $A_I\otimes A_O$. The correspondence with the CP map representation is given by
\begin{equation}
M_{a|x}^{A_IA_O}:=\sum_{jl} \ket{j}\bra{l}^{A_I}\otimes\left[\mm_{a|x}\left(\ket{l}\bra{j}\right)\right]^{A_O^T},
\label{choi}
\end{equation}
where $^T$ denotes transposition in the chosen basis. Complete positivity of the map on the right is equivalent to positive semi-definiteness of the operator on the left. The most general resource for quantum correlations between parties $A, B,\dots$ is represented by a \emph{process matrix} $W^{A_IA_OB_IB_O\dots}$ on the tensor product of input and output spaces~\cite{chiribella08, oreshkov12}. Correlations associated with arbitrary instruments are then given by the generalised Born rule:
\begin{multline}
P(a,b,\dots|x,y,\dots)\\
=\tr\left[\left(M^{A_IA_O}_{a|x} \otimes M^{B_IB_O}_{b|y} \dots \right) 
\cdot W^{A_IA_OB_IB_O\dots}\right].
\label{born}
\end{multline}
This definition subsumes purely spatial resources, which correspond to the special case where all output spaces are trivial. An instrument then reduces to a POVM, namely to a set $\{E_{a|x}\}_a$ of positive operators such that $\sum_a E_{a|x} = \id$, and the process matrix reduces to a density matrix describing a quantum state. A simple example of a temporal resource, on the other hand, is a quantum channel from $A$ to $B$, corresponding to a scenario where $A$ ($B$) has only a non-trivial output (input). Its process-matrix representation is given by the transpose of Eq.~\eqref{choi}:
\begin{equation}
T^{A_OB_I}=\sum_{jl} \ket{j}\bra{l}^{A_O}\otimes\mathcal{T}\left(\ket{j}\bra{l}\right)^{B_I} .
\label{channelchoi}
\end{equation}
The condition that $\mathcal{T}$ is CPTP is equivalent to $T^{A_OB_I}\geq 0$ and $\tr_{B_I} T^{A_OB_I} = \id^{A_O}$. More general spatio-temporal resources are described by process matrices where multiple parties can have non-trivial input and output spaces.

The correspondence of Eq.~\eqref{born} with the usual Born rule allows us to compare temporal and spatial quantum correlations by mapping processes to states, and CP maps to POVM elements---assuming the dimensions of the subsystems in the two scenarios match. Crucially, temporal measurements are described by two Hilbert spaces: input and output. Therefore, the dimension of the corresponding subsystem in a spatial scenario should equal the product of input and output dimensions in the temporal case. For example, a temporal measurement of a qubit is mapped to a spatial measurement of a four-level system in general. (Recall, however, that temporal measurements include preparations and final measurements as particular cases, for which input and, respectively, output dimension is 1.) Such a mapping between an output in the temporal and an input in the spatial scenario can be interpreted as a time reversal operation~\cite{oreshkov2015operational}, although our analysis does not rely on this interpretation. We summarise the relations between spatial and temporal quantum correlations as follows:
\emph{
\begin{enumerate}[(i)]
	\item To every temporal process $W$ corresponds a state $\rho \equiv W/\tr W$; however, there are states that do not have a corresponding process.
	\item \label{measurmentmapping} To every spatial POVM $\{E_{a}\}_a \subset A\cong A_I\otimes A_O$ corresponds a temporal instrument, given by the CP maps $M_a^{A_I A_O} \equiv E_a^{A_I A_O}/d_{A_O}$, where $d_{A_O}$ is the dimension of the output space $A_O$; however, there are instruments that do not correspond to any POVM.
\end{enumerate}
}
Relation (i) arises because all positive semi-definite, normalised density matrices represent a valid quantum state, whereas process matrices have to satisfy additional conditions~\cite{chiribella08, oreshkov12}, e.g. that quantum channels are trace preserving. Relation (ii) follows because, in the Choi representation, CP maps for an instrument sum to a CPTP map, $\tr_{A_O} \sum_a M_a^{A_I A_O} =\id^{A_I}$, while POVM elements must satisfy the stronger requirement $\sum_a E_a^{A} =\id^{A}$.

\paragraph{Resources---} We now use this framework to study spatial and temporal resources, and the operations used to access them. Specifically, we are interested in resources that generate non-trivial statistics (such as correlations) and thus restrict operations to those that do not introduce non-trivial statistics on their own. Such operations must be uncorrelated (represented in the product form of Eq.~\eqref{born}) and must not introduce ``bias'', i.e. cannot deterministically transform a maximally-mixed state into a non-maximally-mixed state. Formally, an instrument $\left\{\mm_{a|x}\right\}_a$ with input (output) dimensions $d^{A_I}$ ($d^{A_O}$) must hence satisfy
\begin{equation}
\sum_a \mm_{a|x}\left(\frac{\id^{A_I}}{d^{A_I}} \right)= \frac{\id^{A_O}}{d^{A_O}}.
\label{nobias}
\end{equation}
In the Choi representation, this condition reads $\tr_{A_I}\sum_a M_{a|x}^{A_IA_O} = \frac{d^{A_I}}{d^{A_O}}\id^{A_O}$. The deterministic preparation of a non-maximally-mixed state would violate this condition and should be understood as a resource, rather than an operation. Conversely, it is easy to verify that projective measurements satisfy Eq.~\eqref{nobias}.

Given the introduced notion of a general quantum resource for generating non-trivial statistics, we now define a criterion to fairly compare quantum and classical resources, independently of their spatial or temporal nature. A general principle advocated for Bell-nonlocality~\cite{woodlesson2012}, non-contextuality~\cite{cavalcanti2017}, and to probe temporal nonclassicality~\cite{Ringbauer2017}, is ``no fine-tuning'': all causal links should manifest in corresponding correlations. Based on this principle, we formulate the following criterion: \emph{Given a quantum (or classical) resource such that, for all free (uncorrelated and unbiased) operations, the generated correlations are no-signalling among certain sets of parties, it should be compared with a classical (or quantum) resource with the same no-signalling constraints.} For details on the definition of classical resources and free operations, see Appendix. Since we consider possible signalling relations directly instead of referring to an underlying causal structure, our criterion differs from no-fine-tuning assumptions in causal modelling~\cite{pearlbook}.

\paragraph{The simplest example---} As a first use case for our framework we now study the simplest instance where temporal correlations arise, which we show to be an interesting special case. Alice and Bob perform temporally-separated measurements on a maximally mixed qubit $\frac{\id^{A_I}}{2}$, which undergoes trivial evolution between the measurements, see Fig.~\ref{fig:Setup}a, as described by the process matrix
\begin{align} \nonumber
W^{A_IA_OB_I} =& \frac{\id^{A_I}}{2}\otimes \Proj{\id}^{A_O B_I} , \\
\Proj{\id}^{X Y} :=& \sum_{jl} \ket{j}\bra{l}^X\otimes \ket{j}\bra{l}^Y .
\label{identityprocess}
\end{align}
The spatial and temporal measurements in this example can be readily identified: For the measurement at $B$, the output is discarded, and an instrument with trivial output space is a POVM. For $A$, an instrument applied on the maximally mixed state is equivalent to one with trivial input, $\rho^{A_O}_{a|x} :=\tr_{A_I} M^{A_IA_O}_{a|x}/2$, corresponding to the preparation of state  $\left(\rho^{A_O}_{a|x} \right)^T /\tr \rho^{A_O}_{a|x}$ with probability $P(a|x)= \tr \rho^{A_O}_{a|x}$. The no-bias condition \eqref{nobias} implies $\sum_a \rho^{A_O}_{a|x} = \id^{A_O}/2$, irrespective of the measurement basis. Therefore, $A$'s instrument can be seen as the temporal correspondent of the spatial POVM $E_{a|x} = 2\rho_{a|x}$. Hence, since all instruments reduce to POVMs, the implication in \emph{(ii)} goes both ways: for every temporal measurement there is a spatial measurement \emph{and} vice versa.

This implies that the correlations Alice and Bob can generate in this temporal scenario are fully equivalent to those obtained in a spatial scenario, where they perform separate measurements on the two-qubit state 
\begin{align}
\ket{\Phi^+}^{A B} = \frac{1}{\sqrt{d}}\sum_j \ket{j}^{A}\otimes \ket{j}^{B} .
\label{pure shared}
\end{align}
This does not imply that the temporal scenario can reproduce the statistics of \emph{any} two-qubit state---for example, a pure product state has no corresponding temporal process---see \emph{(i)}.
A striking consequence, however, is that any joint statistics obtained by $A$ and $B$ in the temporal scenario, $P(a,b|x,y)=\tr\left[ \left(M^{A_IA_O}_{a|x} \otimes E^{B_I}_{b|y}\right) \cdot W^{A_IA_OB_I}\right]$, allow neither signalling from $A$ to $B$ nor from $B$ to $A$, as long as the parties do not have access to additional resources (such as a non-maximally-mixed initial state). Therefore, the above process should be compared to classical resources that do not allow signalling either. This leads to models satisfying ``realism'' and ``no signalling in time''~\cite{brukner2004qet}, which together imply the usual Bell factorisation: $P(a,b|x,y)=\sum_{\lambda} P(a|x,\lambda)P(b|y,\lambda)P(\lambda)$, as we prove in the Appendix. We thus recover full symmetry between the bipartite spatial and temporal scenarios: in both cases, classical correlations cannot violate CHSH inequalities, while quantum correlations can saturate the quantum bound.

\paragraph{Experimental spatio-temporal correlations---} Moving beyond the simplest instance, we now apply our framework to hybrid spatio-temporal scenarios. Consider the tri-partite scenario in Fig.~\ref{fig:Setup}. A measurement is performed on a system qubit which then interacts with a meter qubit followed by spatial measurements on both qubits. Again, $A$ and $B$ measure the system qubit before and after the interaction, respectively. The measurement setting of $C$, which is temporally separated from $A$ but spatially separated from $B$, equates to the basis choice for the meter measurement. As before, we discard $A$'s input system and describe this scenario with a process matrix acting on $A$'s output and $B$ and $C$'s input spaces. We find that this scenario is fully equivalent to a spatial GHZ-state $\ket{GHZ} = \frac{1}{\sqrt{2}}(\ket{000}+\ket{111})$, see Appendix for details. As a consequence, one cannot tell the difference between the two scenarios based on the outcomes of unbiased operations alone.
Experimentally, the strength $\kappa$ of the entangling gate is controlled by initialising the ancilla in the state $\ket{\kappa}=\frac{1}{\sqrt{2}}\left(\sqrt{1+\kappa}\ket{0} + \sqrt{1-\kappa}\ket{1}\right)$. The interaction then generates a family of tripartite entangled states $\rho_{\textsc{abc}}= \ketbra{G_\kappa}_{\textsc{abc}} $,
\begin{equation}
\ket{G_\kappa}_{\textsc{abc}} 
= \id\otimes\id\otimes\frac{\sqrt{1+\kappa}\,\id+\sqrt{1-\kappa}\,\sigma_x}{\sqrt{2}} \ket{GHZ}_{\textsc{abc}},
\label{eq:GHZstate}
\end{equation}
with Pauli matrix $\sigma_x$. The limits are full-strength interaction, $\kappa=1$, and no interaction $\kappa=0$, corresponding to $\ket{\Phi^+}_{\textsc{ab}}\ket{+}_{\textsc{c}}= \frac{1}{2}(\ket{00}+\ket{11})_{\textsc{ab}}(\ket{0}+\ket{1})_{\textsc{c}}$.

\begin{figure}[h!]
  \begin{center}
\includegraphics[width=0.85\columnwidth]{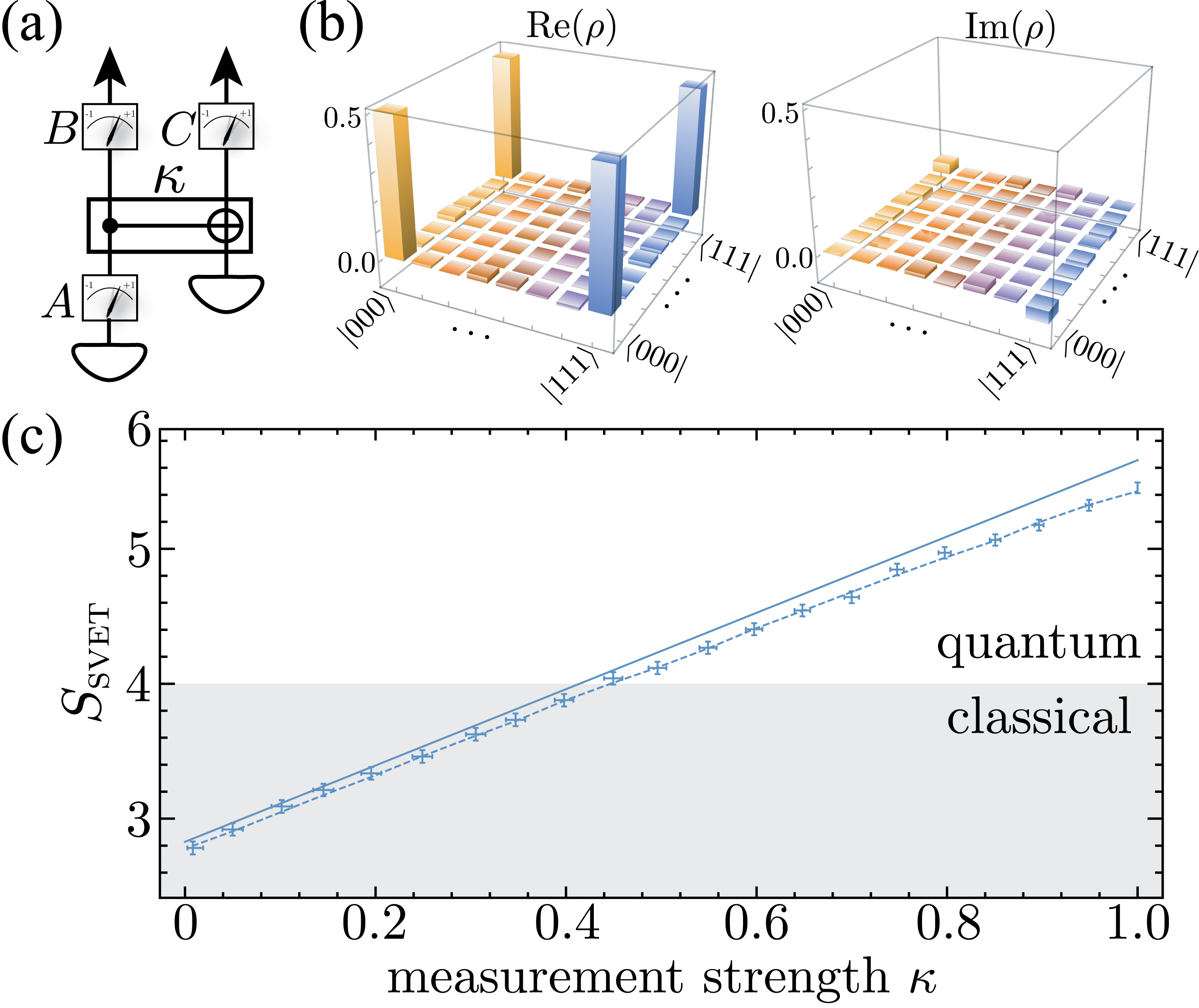}
  \end{center}
\vspace{-1.5em}
\caption{\textbf{Spatio-temporal GHZ state.} \textbf{(a)} Scheme for generating a spatio-temporal GHZ state and \textbf{(b)} the reconstructed density matrix for $\kappa=1$. \textbf{(c)} Violation of the Svetlichny inequality for a range of $\kappa$ witnesses genuine tri-partite entanglement. Error bars for $S_{\textsc{svet}}$ represent $3\sigma$-equivalent statistical confidence regions obtained from $10^5$ MonteCarlo-resampled data points under Poissonian counting statistics; error bars for $\kappa$ correspond to $3\sigma$ confidence intervals estimated from independent classical calibration. The solid line is the theory prediction for perfect states and measurements. The dashed line is the theory prediction taking into account imperfections in the state preparation.}
  \label{fig:ResultsGHZ}
\end{figure}

As before, the restriction to unbiased preparations implies no-signalling between any of the parties. Hence, any non-fined-tuned classical model for this scenario satisfies realism and no-signalling in space and time, implying that all spatial inequalities map onto spatio-temporal inequalities. We now consider the Svetlichny inequality
\begin{align}
S_{\textsc{svet}} = | &\expec{A_0B_0C_0} + \expec{A_0B_0C_1} + \expec{A_0B_1C_0} +\nonumber \\
&\expec{A_1B_0C_0} - \expec{A_1B_1C_0} - \expec{A_1B_0C_1} - \nonumber \\
&\expec{A_0B_1C_1} - \expec{A_1B_1C_1} | \leq 4 .
\label{eq:Svetlichny}
\end{align}
This inequality is satisfied for all bi-separable correlations and thus a sufficient condition for the presence of genuine tri-partite entanglement~\cite{Svetlichny1987,Ajoy2010}, see Appendix for details. In our spatio-temporal scenario we violate inequality~\eqref{eq:Svetlichny} up to $S_{\textsc{svet}}=5.45^{+0.04}_{-0.04}$, demonstrating genuine spatio-temporal tri-partite entanglement, see Fig.~\ref{fig:ResultsGHZ}. All errors correspond to $3\sigma$-equivalent confidence regions obtained from $10^5$ ($5\times 10^3$ for tomography) Monte-Carlo samples based on Poissonian photon counting statistics. On average we recorded ${\sim3000}$ coincidence events per $30$ second measurement. The deviation from theoretical predictions in Fig.~\ref{fig:ResultsGHZ} is due to the limited fidelity $\mathcal{F}=0.964^{+0.002}_{-0.003}$ and purity $\mathcal{P} = 0.942^{+0.04}_{-0.05}$ of the prepared GHZ state; it agrees with predictions adjusted for these imperfections. No-signalling between all parties is satisfied up to a residual variational distance between the observed and uncorrelated distributions of at most $0.02^{+0.03}_{-0.02}$ for $X\to B$. This value is compatible with a no-signalling distribution under Poissonian noise and insufficient to explain the observed violation~\cite{Ringbauer2016Causal}.

\paragraph{Discussion---} Having established a unifying framework for spatio-temporal resources for correlations, we can now address a host of widely unexplored phenomena, including genuine multi-point temporal correlations or measurements on quantum networks~\cite{Chiribella2009} for distributed quantum sensing, temporal decoherence, or temporal steering in quantum communications~\cite{Bartkiewicz2016a}. Particularly intriguing is the question, partially addressed in~\cite{Markiewicz2014,das2013}, how quantum (non-)contextuality relates to no-signalling correlations in space and time. Other directions include extending resource theories for states to general quantum processes~\cite{jia2017generalizing}.
Finally, this improved understanding of how general quantum resources in space or time perform against their classical counterparts will shed new light on the classical simulability of (spatio)temporal correlations~\cite{markiewicz2014genuinely}, which is crucial for applications in quantum communication.

\begin{acknowledgments}
We thank T.~Vulpecula for experimental assistance. This work was supported by the Centres for Engineered Quantum Systems (CE110001013) and for Quantum Computation and Communication Technology (CE110001027) and the UK Engineering and Physical Sciences Research Council (grant number EP/N002962/1). F.C.\ acknowledges support through an ARC DECR Award (DE170100712). M.E.G.\ thanks UQ for support during his sabbatical visit. This publication was made possible through the support of a grant from the John Templeton Foundation (JTF). The opinions expressed in this publication are those of the authors and do not necessarily reflect the views of the JTF. We acknowledge the traditional owners of the land on which UQ is situated, the Turrbal and Jagera people.
\end{acknowledgments}

\newpage
\appendix
\onecolumngrid
\clearpage
\renewcommand{\theequation}{S\arabic{equation}}
\renewcommand{\thefigure}{S\arabic{figure}}
\renewcommand{\thetable}{\Roman{table}}
\renewcommand{\thesection}{S\Roman{section}}
\setcounter{equation}{0}
\setcounter{figure}{0}
\begin{center}
{\bf \large Appendix}
\end{center}
\twocolumngrid

\subsection{Classical temporal resources and operations}
Here we show that bipartite classical temporal resources that do not allow signalling for any free operation cannot be used to violate Bell inequalities.

A temporal classical measurement by a party $A$ is defined by the following variables: a setting variable $x$, an observed outcome $a$, a classical input state $\lambda^A_I$, and an output variable $\lambda^A_O$. Conditioned on the choice of setting $x$, the measurement produces an outcome $a$ and realises a transformation from the input to the output state. The whole operation is thus defined by a conditional probability
\begin{equation}
P_{A}(a,\, \lambda^A_O | x,\, \lambda^A_I), \quad \sum_{a\, \lambda^A_O}P_{A}(a,\, \lambda^A_O | x,\, \lambda^A_I)=1.
\label{classicaloperation}
\end{equation}
Just as for quantum states, a classical operation is considered ``free'', from the perspective of a resource theory for correlations,  if it transforms a maximally mixed state into a maximally mixed state when the outcome is ignored. A classical maximally mixed state is defined as the uniform probability distribution $P(\lambda^A_{I(O)})=\frac{1}{d^A_{I(O)}}$, where $d^A_{I(O)}$ is the number of input (output) states. The ``no bias'' condition for classical operations thus reads 
\begin{equation}
\frac{1}{d^A_{I}}  \sum_{a\, \lambda^A_I}P_{A}(a,\, \lambda^A_O | x,\, \lambda^A_I)
= \frac{1}{d^A_{O}}.
\label{classicalnobias}
\end{equation}
The interpretation is that an operation that does not satisfy \eqref{classicalnobias} requires access to some resource for correlations to be performed. Thus, in a resource theory for correlations, one would separated such additional resources out from the free operations.
 
A classical temporal resource, or classical channel, connecting a party $A$ to a party $B$ is described by a conditional probability distribution 
\begin{equation}
P_{R}(\lambda^B_I|\lambda^A_O), \quad \sum_{\lambda^B_I} P_{R}(\lambda^B_I|\lambda^A_O)  = 1,
\label{classicalresource}
\end{equation}
where $\lambda^B_I$ is Bob's input state and the subscript $R$ stands for ``resource''. By performing free operations on the resource, Alice and Bob can extract correlations from it:
\begin{multline}
P(a,\,b|x,\,y) \\
= \sum_{\vec{\lambda}} P_{B}(b| x,\, \lambda^B_I)  P_{R}(\lambda^B_I|\lambda^A_O) P_{A}(a,\, \lambda^A_O | x),
\label{classicalcorr}
\end{multline}
where $\vec{\lambda}$ denotes the collection of all $\lambda$ variables in the expression. Note that, in this scenario, Bob has trivial output space, so the no bias condition \eqref{classicalnobias} imposes no constraint on his operation $P_{B}(b| x,\, \lambda^B_I)$. On the other hand, Alice has trivial input space, so the no bias condition reads
\begin{equation}
\sum_{a}P_{A}(a,\, \lambda^A_O | x)
= \frac{1}{d^A_{O}}.
\label{alicenobias}
\end{equation}

We can now prove that the correlations \eqref{classicalcorr} satisfy the same factorisation condition as in a local realistic model. To do so, we introduce the two probability distributions
\begin{align}
\bar{P}_R(\lambda^B_I,\, \lambda^A_O) :&= \frac{1}{d^A_O} P_{R}(\lambda^B_I|\lambda^A_O), \\
\bar{P}_{A}(a| x,\, \lambda^A_O ) :&= d^A_O P_{A}(a,\, \lambda^A_O | x).
\label{flipped}
\end{align}
To verify that these are indeed normalised probability distributions, note that Eq.~\eqref{classicalresource} implies $\sum_{\vec{\lambda}} \bar{P}_R(\lambda^B_I,\, \lambda^A_O)=1$, while the no bias condition \eqref{alicenobias} gives $\sum_a \bar{P}_{A}(a| x,\, \lambda^A_O ) =1$. By multiplying and dividing the right-hand-side of Eq.~\eqref{classicalcorr} by $d^A_O$, we can rewrite the correlations in the form
\begin{multline}
P(a,\,b|x,\,y) \\
= \sum_{\vec{\lambda}} P_{B}(b| x,\, \lambda^B_I) \bar{P}_{A}(a| x,\, \lambda^A_O ) \bar{P}_R(\lambda^B_I,\, \lambda^A_O),
\label{Belllocal}
\end{multline}
which manifestly satisfies Bell's local causality condition, and thus cannot violate any Bell inequality.

The above result can be directly extended to a hybrid spatio-temporal resource with an arbitrary number of parties where each party has either trivial input or output space.

\subsection{Tripartite spatio-temporal correlations}
We now consider the tripartite spatio-temporal scenario in Fig.~3 of the main text. Here Alice performs a first measurement on an initially maximally-mixed qubit, which then interacts with an ancillary qubit in the state $\ket{\kappa}=\frac{1}{\sqrt{2}}\left(\sqrt{1+\kappa}\ket{0} + \sqrt{1-\kappa}\ket{1}\right)$ in a controlled-NOT gate. After the interaction, Bob and Charlie perform spatially-separated measurements on the system and ancilla, respectively. Denoting the measurement settings (outcomes) for Alice, Bob and Charlie by $x,y,z$ $(a,b,c)$, respectively, the conditional probabilities are given by
\begin{multline}
P(a,b,c | x,y,z) = \\
\tr\left[\left(M^{A_IA_O}_{a|x} \otimes M^{B_IB_O}_{b|y} \otimes M^{C_IC_O}_{c|z} \right)  \cdot W^{A_IA_OB_IB_OC_IC_O}\right] .
\end{multline}
As discussed in the main text, the initial maximally mixed state can be discarded by replacing the first measurement with $\rho^{A_O}_{a|x} :=\tr_{A_I} M^{A_IA_O}_{a|x}/2$, corresponding to the preparation of state  $\left(\rho^{A_O}_{a|x} \right)^T /\tr \rho^{A_O}_{a|x}$ with probability $P(a|x)= \tr \rho^{A_O}_{a|x}$. 
Similarly, the final measurements simply correspond to the POVM elements $E^{B_I}_{b|y} :=\tr_{B_O} M^{B_IB_O}_{b|y}/2$ for Bob and $E^{C_I}_{c|z} :=\tr_{C_O} M^{C_IC_O}_{c|z}/2$ for Charlie.
The process matrix now takes the form
\begin{multline}
 W^{A_IA_OB_IB_OC_IC_O} = \chi_\kappa^{A_OB_IC_I} \\ 
=\tr_D[(\id^{A_O}\otimes \ketbra{\kappa}^{D^T} \otimes  \id^{B_I} \id^{C_I})\Lambda_{\textsc{cnot}}],
\end{multline}
where the partial trace is over the ancillary qubit (denoted $D$) and the Choi matrix of the CNOT gate, $\lambda_{\textsc{cnot}}$ is, following Eq.~(4) of the main text,
\begin{align}
\Lambda_{\textsc{cnot}} &=  \ketbra{\lambda_{\textsc{cnot}}},\\ \nonumber
\ket{\lambda_{\textsc{cnot}}} &=\sqrt{2}\left(\ket{0 0}^{A_O B_I}\ket{\Phi^+}^{D C_I} + \ket{1 1}^{ A_O B_I}\ket{\Psi^+}^{D C_I}\right),
\label{cnotchoi}
\end{align}
with $\ket{\Psi^+}=\left(\ket{01}+\ket{10}\right)/\sqrt{2}$. A straightforward calculation shows that $\chi_\kappa^{A_OB_IC_I}=2 \rho_{\textsc{abc}} = 2 \ketbra{G_k}_{\textsc{abc}}$, given by Eq.~(8) of the main text. 

Hence, comparing with the usual Born rule for spatially separated measurements on three qubits in the state $\rho_{\textsc{abc}}$ establishes the equivalence
\begin{equation}
\!\!\tilde P(a,b,c | x,y,z) = \tr\left[\left(E^{A}_{a|x} \otimes E^{B}_{b|y} \otimes E^{C}_{c|z} \right)  \cdot \rho_{\textsc{abc}} \right],
\end{equation}
where Alice's state preparation is mapped to the measurement of $E^{A}_{a|x} = 2 \rho^{A}_{a|x}$.

As noted in the main text, the restriction to unbiased preparation implies no signalling between any of the parties. Hence, any faithful (i.e.\ not fined-tuned) classical model for this scenario satisfies realism and no-signalling in space and time, which in turn imply that all spatial Bell-type inequalities map onto spatio-temporal inequalities.

Specifically, when choosing measurement settings $A_0=B_0=C_0=\hat X$ and $A_1=B_1=C_1=\hat Y$ the GHZ-state $\ket{GHZ}$ gives rise to the so-called \emph{GHZ-paradox}~\cite{Greenberger1990}, which refers to a classically impossible assignment of measurement outcomes
\begin{equation}
\begin{split}
A_0B_0C_0 &=  1 \\
A_0B_1C_1&= -1 \\
A_1B_0C_1&= -1 \\
A_1B_1C_0&= -1 .
\end{split}
\label{eq:GHZparadox}
\end{equation}
Although classically impossible, this assignment can ideally be satisfied by a perfect GHZ state. In practice, following Mermin~\cite{Mermin1990a}, the GHZ paradox can be made experimentally robust, by phrasing it in terms of an inequality
\begin{align}
S_{\textsc{ghz}} &= \expec{A_0B_0C_0} - \expec{A_0B_1C_1} - \nonumber\\
&\quad \expec{A_1B_0C_1} - \expec{A_1B_1C_0} \leq 2
\label{eq:MerminInequality}
\end{align}
A violation of Eq.~\eqref{eq:MerminInequality} can be used to demonstrate entanglement between $A$, $B$, and $C$. However, it is not sufficient to demonstrate genuine tri-partite entanglement. In other words, correlations violating Eq.~\eqref{eq:MerminInequality} may originate from a bipartite entangled state with the third qubit merely classically correlated to the others. 
To overcome this, Svetlichny considered a scenario where two of the parties have to violate one of two complementary CHSH inequalities, depending on the third party's input. This task can only be accomplished with genuine tri-partite entanglement and the Svetlichny inequality is thus satisfied for any bi-separable state~\cite{Svetlichny1987,Ajoy2010}
\begin{align}
S_{\textsc{svet}} = | &\expec{A_0B_0C_0} + \expec{A_0B_0C_1} + \expec{A_0B_1C_0} +\nonumber \\
&\expec{A_1B_0C_0} - \expec{A_1B_1C_0} - \expec{A_1B_0C_1} - \nonumber \\
&\expec{A_0B_1C_1} - \expec{A_1B_1C_1} | \leq 4 .
\end{align}
Note that the inequality is symmetric under exchange of parties. It can be violated up to a value of $S_{\textsc{svet}}=4\sqrt{2}$ using measurements in the \textsc{xy}-plane of the Bloch sphere. Denoting $A_0/B_0/C_0 = \cos \phi_{\textsc{a/b/c}} \hat X + \sin \phi_{\textsc{a/b/c}} \hat Y$, $A_1/B_1/C_1 = -\sin \phi_{\textsc{a/b/c}} \hat X + \cos \phi_{\textsc{a/b/c}} \hat Y$ the measurement settings correspond to $\phi_{\textsc{a}}=\pi/4$, $\phi_{\textsc{b}}=0$, and $\phi_{\textsc{c}}=\pi/2$. Note, that the violation of the Svetlichny inequality is a sufficient, but not necessary condition for genuine tri-partite entanglement~\cite{Brunner2014Review}.

\end{document}